\documentclass[aps,pra,letterpaper,10pt,twocolumn,showpacs,superscriptaddress]{revtex4-1}
\usepackage{hyperref}
\usepackage{amssymb}
\usepackage{epsfig}
\usepackage{amsmath}
\usepackage{graphicx}
\usepackage{subfigure}
\usepackage{amsfonts}
\usepackage{epstopdf}

\newcommand{\rmB}{\mathrm{B}}

\newcommand{\rmm}{\mathrm{m}}
\newcommand{\rmb}{\mathrm{b}}
\newcommand{\rmc}{\mathrm{c}}

\newcommand{\rmw}{\mathrm{w}}
\newcommand{\rmin}{\mathrm{in}}
\newcommand{\rmout}{\mathrm{out}}

\newcommand{\trp}{\mathsf{T}}
\newcommand{\calE}{\mathcal{E}}

\begin{document}
\title{Entangling two distant non-interacting microwave modes}

\author{M. Abdi}
\affiliation{Technische Universit\"at M\"unchen, Physik Department, James Franck Str., D-85748 Garching, Germany}

\author{P. Tombesi}
\author{D. Vitali}
\affiliation{School of Science and Technology, Physics Division, University of Camerino, via Madonna delle Carceri, 9, I-62032 Camerino (MC), Italy, and INFN, Sezione di Perugia, Italy}

\date{\today}
\begin{abstract}
We propose a protocol able to prepare two remote and initially uncorrelated microwave modes in an entangled stationary state, which is certifiable using only local optical homodyne measurements.
The protocol is an extension of continuous variable entanglement swapping, and exploits two hybrid quadripartite opto-electro-mechanical systems in which a nanomechanical resonator acts as a quantum interface able to entangle optical and microwave fields.
The proposed protocol allows to circumvent the problems associated with the fragility of microwave photons with respect to thermal noise and may represent a fundamental tool for the realization of quantum networks connecting distant solid-state and superconducting qubits, which are typically manipulated with microwave fields. The certifying measurements on the optical modes guarantee the success of entanglement swapping without the need of performing explicit measurements on the distant microwave fields.
\end{abstract}

\maketitle

%
%
\section{Introduction}

Quantum information processing based on superconducting qubits is rapidly becoming a promising avenue for the implementation of quantum computation tasks~\cite{Lucero2012,Reed2012,Barends2014}. In fact these qubits can be easily manipulated and controlled by microwave fields through transmission line resonators, and various examples of quantum states of microwave and array of qubits have been demonstrated~\cite{Riste2013,Kirchmair2013}. A limitation associated with microwave fields is due to their low frequency, which makes them fragile with respect to thermal noise, and forces one to operate in dilution refrigerator environments. As a consequence, it is not easy to transfer quantum states of microwave fields over long distances, and their direct use in a quantum information network formed by distant nodes seems prohibitive.

Here we propose a scheme for the implementation of a fundamental tool for the realization of quantum networks of solid state/superconducting qubits, i.e., the generation of robust entanglement between two distant microwave fields. The scheme is an extension of continuous variable (CV) entanglement swapping and its key ingredient is the ability of nanomechanical resonators of acting as quantum interfaces between optical and microwave fields~\cite{Tian2010,Regal2011,Barzanjeh2011,Barzanjeh2012,Taylor2011,Wang2012,Tian2012}. In fact, a straightforward solution for entangling two distant microwave modes---and potentially two nodes of a quantum information network, is to use a standard entanglement swapping protocol~\cite{Pan1998,Jia2004,Takei2005,Pirandola2006} between two distant nodes. The simplest option is to prepare two entangled microwave modes in each site and send one mode for each site to an intermediate site for a Bell measurement, which in the CV case amounts to two joint homodyne detections after mixing the two fields on a balanced beam-splitter. Differently from the optical case, in the microwave case the protocol is inefficient because the quantum state of the microwave fields is seriously degraded by their propagation over long distances, due to their unavoidable thermalization. The detrimental effect of thermal noise on microwave photons also seriously hinders the experimental \emph{verification} of the presence of long-distance microwave entanglement. In fact, this is usually achieved by performing joint correlated homodyne measurements on the two distant fields, which also requires propagating the microwave fields over long distances.

In the present proposal we solve both problems by adopting a scheme which combines in a non-trivial way two different ingredients which have been recently proposed in the literature: i) the generation of CV optical-microwave entanglement by means of the common interaction of these modes with a nanomechanical resonator interfacing them at the quantum level~\cite{Barzanjeh2012}; ii) the protocol of entanglement swapping with local certification of Ref.~\cite{Abdi2012}, able to warrant the presence of long-distance entanglement even without performing direct measurements on the remote nodes.
In such a scheme each node possesses a \emph{quadripartite} opto-electro-mechanical system in which a nanomechanical resonator interface is coupled to one microwave mode and to \emph{two} optical modes, generating a robust stationary state in which bipartite entanglement between the microwave and an optical mode, and between the two optical modes is simultaneously present. The two optical modes at each site, much less affected by thermal noise, can be then safely sent to the intermediate site where a first pair is subject to the entangling Bell measurement, and the second pair is subject to a set of homodyne measurements allowing to certify locally the presence of entanglement between the microwave modes at the two remote sites (see Fig.~1). We will show that such a quadripartite opto-electro-mechanical system can be prepared in the required class of ``certifying'' states and that the proposed entangling protocol can be successfully demonstrated using currently available technology.

In Sec.~II we recall the basic ingredients of the protocol of entanglement swapping with local certification of Refs.~\cite{Abdi2012,Abdi2014}, while in Sec.~III we determine the conditions under which the two opto-electro-mechanical systems are able to implement it. In Sec.~IV we describe the resulting Gaussian stationary state which is the basic resource for the realization of entanglement with local certification, in Sec.~V we present numerical results in the case of a feasible experimental scenario, and Sec~VI is for concluding remarks.

\begin{figure}[tbs]
\includegraphics[width=3.0 in]{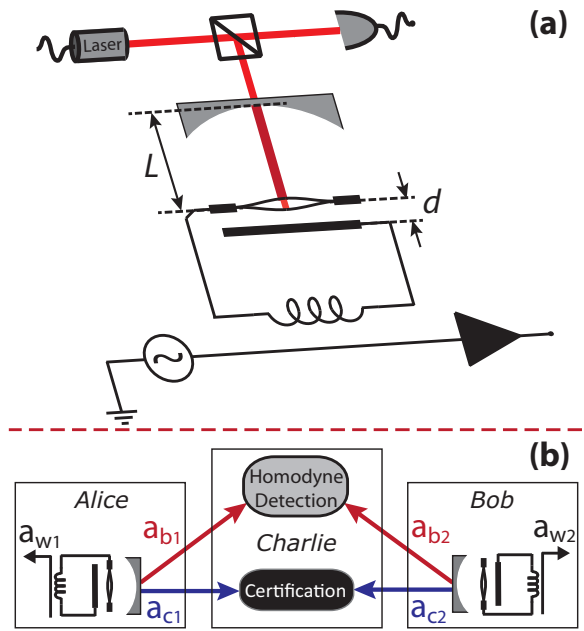}
\caption{(a) A possible scheme for the opto-electro-mechanical system: a lumped-element microwave cavity is capacitively coupled to a mechanical resonator which is also coupled to an optical cavity formed by an input mirror and the optically coated drum-head capacitor.
(b) Scheme for the entanglement swapping protocol.}
\label{scheme}
\end{figure}

%
%
\section{The basics of the protocol}
In order to create an entangled state of two initially uncorrelated distant microwave modes, we employ entanglement swapping~\cite{Pan1998,Jia2004,Takei2005,Pirandola2006}, supplemented with the local certification protocol proposed in~\cite{Abdi2012,Abdi2014}, and apply it to the case when the two remote sites Alice and Bob possess each a continuous variable (CV) \emph{quadripartite} opto-electro-mechanical system formed by a mechanical resonator simultaneously interacting with two optical cavity modes and a microwave cavity mode. Such a system is an extension of the tripartite opto-electro-mechanical systems theoretically studied in Refs.~\cite{Tian2010,Regal2011,Barzanjeh2011,Barzanjeh2012,Taylor2011,Wang2012,Tian2012} (see Fig.~\ref{scheme}a), and whose first experimental realizations have been reported in Refs.~\cite{Winger2011,Bochmann2013,Bagci2014,Andrews2014}. In such systems, a nanomechanical resonator with high mechanical quality factor is simultaneously coupled to a microwave/radiofrequency mode and to an optical mode, acting therefore as an interface between the two fields at completely different wavelengths.

The basic ingredient for creating a quantum link between the two distant microwave fields is that the state at each remote node possesses a nonzero entanglement between the microwave mode and a traveling optical mode. In fact Alice and Bob, profiting from the fact that traveling optical field are much less affected by decoherence than microwave/rf fields, can each send his/her optical mode to a third node halfway between them (Charlie, see Fig.~1b). Charlie can then perform a Bell measurement on the two optical modes and consequently entangle the two distant microwave modes by means of CV entanglement swapping. Here we supplement the protocol with the \emph{local certification} of Refs.~\cite{Abdi2012,Abdi2014}, which allows Charlie also to \emph{certify} that the two remote modes have been successfully entangled, without explicitly performing joint correlated measurements on the two distant CV systems. Charlie receives the two optical modes from Alice and Bob, then measures a first pair for the Bell measurement, and subsequently the second pair for certifying the presence of entanglement between the remote nodes. The local certification is obtained when the \emph{tripartite} systems formed by the two optical modes and the microwave mode at Alice and Bob sites are in a state satisfying an appropriate \emph{certifying} condition.

In order to be more specific, we assume a symmetric situation, i.e., Alice and Bob initially possess the \emph{same} CV tripartite state. We can also always restrict ourselves to the case when this state is a zero-mean Gaussian state, that is, fully determined by its Gaussian characteristic function $\Phi(\boldsymbol{k})=\exp(-\frac{1}{2}\boldsymbol{k}^{\trp}\mathbf{Vk})$, where $\boldsymbol{k}\in\mathbb{R}^{6}$ is a vector of real variables associated with the CV of the three modes, and $\mathbf{V}$ is the $6 \times 6$ covariance matrix (CM), which can be written in the following block form
\begin{equation}
\mathbf{V}=\left[
\begin{array}[c]{ccc}
\mathbf{W} & \mathbf{D} & \mathbf{F}\\
\mathbf{D}^{\trp} & \mathbf{B} & \mathbf{E}\\
\mathbf{F}^{\trp} & \mathbf{E}^{\trp} & \mathbf{C}%
\end{array}
\right],
\label{CM}
\end{equation}
where the blocks $\mathbf{W,B,C,D,E,F}$ are $2\times2$ submatrices.
The two following submatrices can be extracted from the above CM
\begin{equation}
\mathbf{V}_{\rmw\rmb}=\left[
\begin{array}[c]{cc}
\mathbf{W} & \mathbf{D}\\
\mathbf{D}^{\trp} & \mathbf{B}
\end{array}\right],~
\mathbf{V}_{\rmb\rmc}=\left[
\begin{array}[c]{cc}
\mathbf{B} & \mathbf{E}\\
\mathbf{E}^{\trp} & \mathbf{C}
\end{array}\right],
\end{equation}
where $\mathbf{V}_{\rmw\rmb}$ describes the remote-Bell modes, and $\mathbf{V}_{\rmb\rmc}$ the Bell-certification modes.

First, Alice and Bob keep one mode (a microwave mode in our case) and send the other two modes (the Bell and certification optical modes) to Charlie (see Fig.~\ref{scheme}b).
In the next step, Charlie performs a CV Bell measurement on the optical Bell modes, by using a balanced beam splitter and two homodyne detectors~\cite{Pirandola2006}.
In this way, two combinations of field quadratures are measured and the outcomes of the measurement are stored by Charlie. The resulting random displacement of the state due to the Bell measurement can be eliminated by means of optimal displacements with suitable gains~\cite{Hoelscher-Obermaier2011,Abdi2014}, and the output state of the remaining remote and certifying modes is a quadripartite Gaussian state with CM of the form
\begin{equation}
\mathbf{V}_{\mathrm{out}}=\left[
\begin{array}[c]{cc}
\mathbf{V}_{\rmw1,\rmw2} & \mathbf{X}\\
\mathbf{X}^{\trp} & \mathbf{V}_{\rmc1,\rmc2}
\end{array}\right],
\label{outcm}
\end{equation}
where $\mathbf{V}_{\rmw1,\rmw2}$ describes the bipartite subsystem composed of Alice and Bob's remote (microwave) modes, while $\mathbf{V}_{\rmc1,\rmc2}$ describes the certification modes at Charlie's site.
The cross-correlation elements are included in $\mathbf{X}$.

We quantify bipartite entanglement in terms of the logarithmic negativity which is defined as $E_{N}=\mathrm{max}\{0,-\log(2\eta_{-})\}$, where $\eta_{-}$ is the minimum symplectic eigenvalue of the partial transposed two-mode CM~\cite{Eisert2001,Vidal2002,Plenio2005}. The minimum symplectic eigenvalues of the partially-transposed $4\times 4$ matrices, $\mathbf{V}_{\rmw1,\rmw2}$ and $\mathbf{V}_{\rmc1,\rmc2}$ can be computed and expressed only in terms of state purities~\cite{Pirandola2006}, $\mu=\mathrm{Tr}\{\rho^{2}\}$, and which for a $N$-mode Gaussian state are given by $\mu_{N}=(2^{N}\sqrt{\det\mathbf{V}_{N}})^{-1}$, where $ \mathbf{V}_{N}$ is the CM of the state.
By expressing the CMs in their standard form~\cite{Wang2003,Adesso2006} one arrives at:
\begin{equation}
\eta_{-}(\mathbf{V}_{\rmw1,\rmw2}^{\mathsf{PT}})=\frac{\mu_{\rmb}}{2\mu_{\rmw\rmb}},\;\;\;\;\;\;
\eta_{-}(\mathbf{V}_{\rmc1,\rmc2}^{\mathsf{PT}})=\frac{\mu_{\rmb}}{2\mu_{\rmb\rmc}}. \label{sympl-relat}
\end{equation}
The entanglement swapping protocol is successful and the establishment of entanglement between the remote Alice and Bob sites can be certified by Charlie when
\begin{equation}
E_{N}^{\rmw1,\rmw2} > E_{N}^{\rmc1,\rmc2} > 0,
\label{fund}
\end{equation}
because in this case detection of any entanglement by Charlie in the certification modes guarantees the generation of entanglement between the remote modes.
Eqs.~(\ref{sympl-relat}) show that this condition is satisfied if the local and global purities of the initial tripartite Gaussian state satisfies the \emph{certifying condition} of Refs.~\cite{Abdi2012,Abdi2014}
\begin{equation}
\mu_{\rmw\rmb} > \mu_{\rmb\rmc} > \mu_{\rmb}.
\label{tcs}
\end{equation}

%
%
\section{The opto-electro-mechanical system at each site}

As shown in Refs.~\cite{Abdi2012,Abdi2014}, the certifying condition of Eq.~\ref{tcs}) implies that both the microwave-optical Bell $wb$ bipartite subsystem and the optical Bell-certifying $bc$ bipartite subsystem are entangled. However, entangling microwaves and optical fields is not trivial at all due to the completely different wavelengths, and in fact, it has not been experimentally demonstrated yet. However, as suggested in \cite{Regal2011,Barzanjeh2011,Barzanjeh2012}, a promising solution is provided by opto-electro-mechanical systems, in which a nanomechanical resonator is simultaneously coupled to an optical and a microwave cavity mode. In particular, Refs.~\cite{Barzanjeh2011,Barzanjeh2012} show that if the optical and the microwave modes are driven on opposite sidebands (i.e., one on the red and on on the blue sideband) one has an effective parametric amplifier with an optical idler (signal) and a microwave signal (idler).
Therefore a viable scheme is to provide both Alice and Bob with a \emph{quadripartite} CV opto-electro-mechanical system formed by a nanomechanical resonator which is coupled to a driven microwave cavity mode and \emph{two} optical modes. As shown in the previous Section, the two distant microwave modes can be entangled (and this fact can be locally certified by Charlie) if the tripartite reduced state at each site obtained by tracing out the mechanical resonator is a Gaussian certifying state.

The two traveling optical modes which both Alice and Bob have to send to Charlie could be obtained by driving a single optical cavity mode, and then extracting two independent output optical modes by suitably filtering the
outgoing field as in~\cite{Genes2008}. However, it is more
efficient to drive \emph{two different} cavity modes and filtering
one output mode \cite{Giovannetti2001a,Genes2009} for each driven mode, and we
shall consider this latter situation from now on.

The Hamiltonian of the desired opto-electro-mechanical systems located at each site is therefore the sum of an optical, microwave and mechanical term, $\hat{H}=\hat{H}_{\mathrm{oc}}+\hat{H}_{\mathrm{mw}}+\hat{H}_{\rmm}$, where
\begin{subequations}
\begin{align}
\hat{H}_{\mathrm{oc}}&=\hbar\omega_{\rmb}(\hat{q}) ~\hat{a}_{\rmb}^{\dagger}\hat{a}_{\rmb} +\hbar\omega_{\rmc}(\hat{q}) ~\hat{a}_{\rmc}^{\dagger}\hat{a}_{\rmc}, \\
\hat{H}_{\mathrm{mw}}&=\hbar\omega_{\rmw}(\hat{q}) ~\hat{a}_{\rmw}^{\dagger}\hat{a}_{\rmw}, \\
\hat{H}_{\rmm}&=\frac{\hbar\omega_{\rmm}}{2}(\hat{p}^{2}+\hat{q}^{2}),
\label{hamiltonian}
\end{align}
\end{subequations}
with $\hat{a}_{x}$ and $\hat{a}_{x}^{\dagger}$, $x = \rmb, \rmc, \rmw$, the annihilation and creation operators of the Bell optical cavity mode, of the certification optical cavity mode, and of the microwave cavity mode respectively.
The mechanical oscillator with mass $m$ and natural frequency $\omega_{\rmm}$ is described by dimensionless momentum and position operators $\hat{p}$ and $\hat{q}$ ($[\hat{q},\hat{p}]=i$).
All cavity modes interact with the mechanical resonator through the dispersive parametric coupling due to the dependence of the cavity mode frequencies upon the effective resonator position $\hat{q}$, $\omega_{x}(\hat{q})$. Here we consider the most typical case in which one can safely describe this dependence at first order in $\hat{q}$, and assume $\omega_{x}(\hat{q})\simeq \omega_{x}-g_x \hat{q}$, even though there are cases where also the quadratic term is responsible for appreciable effects~\cite{Thompson2008,Rocheleau2010,Sankey2010,Karuza2012}. The parameters $\omega_{x}$ and $g_x$ are the unperturbed cavity mode frequencies and coupling rates respectively.

Therefore, by including the external drives, the Hamiltonian of the quadripartite linearly coupled opto-electro-mechanical system takes the following form
\begin{align}\label{eq:hoem}
\hat{H}_{\mathrm{OEM}} =&\frac{\hbar\omega_{\rmm}}{2}(\hat{p}^{2}+\hat{q}^{2})+\sum_{x=\rmb,\rmc,\rmw}\hbar(\omega_{x}-g_{x} \hat{q})\hat{a}_{x}^{\dagger}\hat{a}_{x} \\
&-i\sum_{x=\rmb,\rmc,\rmw}\hbar \calE_{x}\big(\hat{a}_{x}e^{i\omega_{0x}t}-\hat{a}_{x}^{\dagger}e^{-i\omega_{0x}t}\big). \nonumber
\end{align}
In the driving terms $\omega_{0x}$ are the frequencies of the driving fields, while the driving rates $\calE_{x}$ at which optical and microwave photons are pumped within the cavities depend upon the input powers $P_{x}$ and the cavity damping rates through the input ports $\kappa_{x}$ according to $\calE_{x}= \sqrt{2P_{x}\kappa_{x}/\hbar\omega_{0x}}$.

\subsection*{Dynamics}
The dynamics of the opto-electro-mechanical system at each site can be described in terms of quantum Langevin equations (QLE), i.e., by the Heisenberg equations associated with the system Hamiltonian $\hat{H}_{\mathrm{OEM}}$ of Eq.~(\ref{eq:hoem}) supplemented with damping and noise terms caused by the interaction of each mode with the respective optical, microwave, and mechanical reservoirs~\cite{Hammerer2012}.
The resulting Langevin equations are nonlinear, but we are interested in the regime where both the optical and the microwave modes are intensely driven so that all operators can be expressed in terms of a large semiclassical value plus a fluctuation operator, $\hat{a}_{x} \to \langle a_{x}\rangle +\hat{a}_{x}$. In fact, the single-photon opto- or electro-mechanical couplings are too weak to create any entanglement. Thus they are enhanced by strong drives. The steady state semiclassical values are then given by
\begin{align}
\langle a_{x}\rangle &= \frac{\calE_{x}}{\kappa_{x}+i\Delta_{x}}, \;\;x = \rmb, \rmc, \rmw\nonumber\\
\langle q\rangle &= \frac{\sum_{x}g_{x}|\langle{a}_{x}\rangle|^{2}}{\omega_{\rmm}}, \\
\langle p\rangle &= 0, \nonumber
\end{align}
where $\Delta_{x}\equiv \omega_{x}-\omega_{0x} -g_{x} \langle q\rangle$ are the detunings of optical and microwave modes. If the number of optical and microwaves photons within the cavity are high enough, i.e., the intracavity amplitudes satisfy $|\langle a_{x}\rangle| \gg 1$ one can safely neglect the nonlinear terms in the equation of motion for the fluctuation operators and consider the \emph{linearized} QLEs
\begin{subequations}
\begin{eqnarray}
\dot{\hat{q}}&=& \omega_{\rmm}\hat{p}, \\
\dot{\hat{p}}&=& -\omega_{\rmm}\hat{q} -\gamma_{\rmm}\hat{p} +\sum_{x} g_{x}\langle a_{x}\rangle(\hat{a}_{x}^{\dagger} +\hat{a}_{x}) +\hat{\xi}, \\
\dot{\hat{a}}_{x}&=& -(\kappa_{x}+i\Delta_{x})\hat{a}_{x} +ig_{x}\langle a_{x}\rangle\hat{q} +\sqrt{2\kappa_{x}}~\hat{a}_{x}^{\rmin},
\end{eqnarray}\label{langevin}
\end{subequations}
where $x = \rmb, \rmc, \rmw$. $\gamma_\mathrm{m}$ is the rate at which mechanical energy is damped to the environment, while $\hat{\xi}(t)$ is a Brownian stochastic force with zero mean value whose correlation function can be well approximated by~\cite{Giovannetti2001}
\begin{equation}
\label{eq:browncorre2}\bigl\langle \hat{\xi}(t)\hat{\xi}(t^{\prime})\bigr\rangle \approx \gamma_\mathrm{m}\left[(2\bar{n}+1) \delta(t-t^{\prime})+\mathrm{i}
\frac{\delta^{\prime}(t-t^{\prime})}{\omega_\mathrm{m}}\right]  ,
\end{equation}
where $k_{\mathrm{B}}$ is the Boltzmann constant, $\bar{n}=\left[\exp (\hbar \omega _\mathrm{m}/k_{\mathrm{B}}T)-1\right] ^{-1} \approx k_{\mathrm{B}} T/\hbar \omega_\mathrm{m}$ is the mean thermal phonon number at temperature $T$, and $\delta^{\prime}(t-t^{\prime})$ denotes the derivative of the Dirac delta.

The optical and microwave modes are instead affected by the input noises $\hat{a}_{x}^{\rmin}(t)$, whose correlation functions are given by~\cite{Gardiner2000}
\begin{subequations}
\begin{eqnarray}
\big\langle\hat{a}_{x}^{\rmin}(t)\hat{a}_{x'}^{\rmin,\dagger}(t')\big\rangle &=&\delta_{x,x'}[\bar{n}(\omega_{x})+1]\delta(t-t'), \\
\big\langle\hat{a}_{x}^{\rmin,\dagger}(t)\hat{a}_{x'}^{\rmin}(t')\big\rangle &=&\delta_{x,x}\bar{n}(\omega_{x})\delta(t-t'),
\end{eqnarray}
\label{noise}
\end{subequations}
where $\bar{n}(\omega)\equiv [\exp(\hbar\omega/k_{\rmB}T)-1]^{-1}$ is the mean thermal photon number. For the optical modes it is $\bar{n}(\omega_{\rmb}) \simeq \bar{n}(\omega_{\rmc})\simeq 0$, while thermal excitations cannot be neglected in the microwave case, even at cryogenic temperatures, due to the much lower frequency, and this is the reason why microwave fields are much more sensitive to thermal noise.

%
%
\section{The opto-electro-mechanical Gaussian state at each site}

We now see how the tripartite Gaussian state for two optical modes and one microwave mode which is needed for implementing the protocol of entanglement swapping with local certification emerges from the dynamics described above. Introducing optical and microwave quadrature field operators defined by $\hat{a}_{x}\equiv (\hat{X}_{x}+i\hat{Y}_{x})/\sqrt{2}$, the linearized QLE of Eqs.~(\ref{langevin}) can be rewritten in the following compact form
\begin{equation}
\dot{\hat{\boldsymbol{u}}}=\mathbf{A}\hat{\boldsymbol{u}} +\hat{\boldsymbol{n}},
\label{compact}
\end{equation}
where the vector of the system operators and the vector of the system noises are respectively defined as
\begin{align*}
\hat{\boldsymbol{u}} &\equiv [\hat{q},\hat{p},\hat{X}_{\rmb},\hat{Y}_{\rmb},\hat{X}_{\rmc},\hat{Y}_{\rmc},\hat{X}_{\rmw},\hat{Y}_{\rmw}]^{\trp}, \\
\hat{\boldsymbol{n}} &\equiv [0,\hat{\xi},\sqrt{2\kappa_{\rmb}}\hat{X}_{\rmb}^{\rmin},\sqrt{2\kappa_{\rmb}}\hat{Y}_{\rmb}^{\rmin},\sqrt{2\kappa_{\rmc}}\hat{X}_{\rmc}^{\rmin}, \\
&~~~\times\sqrt{2\kappa_{\rmc}}\hat{Y}_{\rmc}^{\rmin},\sqrt{2\kappa_{\rmw}}\hat{X}_{\rmw}^{\rmin},\sqrt{2\kappa_{\rmw}}\hat{Y}_{\rmw}^{\rmin}]^{\trp},
\end{align*}
while $\mathbf{A}$ is the drift matrix of the system, given by
\begin{widetext}
\begin{equation}
\mathbf{A}\equiv \left[
\begin{array}{cccccccc}
0 & \omega_{\rmm} & 0 & 0 & 0 & 0 & 0 & 0 \\
-\omega_{\rmm} & -\gamma_{\rmm} & G_{\rmb} & 0 & G_{\rmc} & 0 & G_{\rmw} & 0 \\
0 & 0 & -\kappa_{\rmb} & \Delta_{\rmb} & 0 & 0 & 0 & 0 \\
G_{\rmb} & 0 & -\Delta_{\rmb} & -\kappa_{\rmb} & 0 & 0 & 0 & 0 \\
0 & 0 & 0 & 0 & -\kappa_{\rmc} & \Delta_{\rmc} & 0 & 0 \\
G_{\rmc} & 0 & 0 & 0 & -\Delta_{\rmc} & -\kappa_{\rmc} & 0 & 0 \\
0 & 0 & 0 & 0 & 0 & 0 & -\kappa_{\rmw} & \Delta_{\rmw} \\
G_{\rmw} & 0 & 0 & 0 & 0 & 0 & -\Delta_{\rmw} & -\kappa_{\rmw}
\end{array}\right].
\end{equation}
\end{widetext}
Here $G_{x} =\sqrt{2}\langle a_{x}\rangle g_{x}$ are the effective opto- and electro-mechanical couplings which are enhanced by the steady state intracavity optical and microwave steady state amplitudes.

The opto-electro-mechanical system is stable and reaches a steady state after a transient time if all the eigenvalues of the drift matrix $\mathbf{A}$ have negative real part, i.e., it satisfies stability conditions~\cite{Ogata2010}.
Since all noise terms in Eq.~(\ref{compact}) are zero-mean Gaussian and the dynamics are linear, this steady state for the mechanical and cavity mode fluctuations is a quadripartite zero-mean Gaussian state, fully determined by its $8\times 8$ CM. However, in order to implement the entanglement swapping protocol we need to use \emph{traveling} optical modes at the output of the cavities rather than intracavity modes. Using the standard input-output relation for the cavity fields~\cite{Gardiner2000}, $\hat{a}_{x}^{\rmout}=\sqrt{2\kappa_{x}}\hat{a}_{x} -\hat{a}_{x}^{\rmin}$, and the approach of Ref.~\cite{Genes2008}, one can define the selected output modes by means of the bosonic annihilation operators $\hat{a}_{x}^{\mathrm{sel}}=\int_{t_{0}}^{t}h_{x}(t-s)\hat{a}_{x}^{\rmout}(s)ds$.
The causal filter function $h_{x}(t)$ defines the output mode, is characterized by a central frequency and a bandwidth, and must be normalized to one in order to ensure that the bosonic commutators still hold.
We choose for simplicity the same form of filter function for both optical and microwave fields, that is, $h_{x}(t)=\sqrt{2/\tau_{x}}\Theta(t)\exp[(-1/\tau_{x} +i\Omega_{x})t]$ where $\Theta(t)$ is the Heaviside step function, $1/\tau_{x}$ is bandwidth of the filter and $\Omega_{x}$ is its central frequency.

The tripartite system of interest for the application of the modified entanglement swapping protocol proposed in Sec.~II is therefore the one formed by the three cavity output fields (two optical and one microwave field) defined by the filtered annihilation operators $\hat{a}_{x}^{\mathrm{sel}}$, and obtained by tracing out the mechanical resonator interfacing the optical and microwave modes. At the steady state, this system is in Gaussian CV state, fully characterized by its symmetrically ordered $6\times 6$ CM whose elements are
\begin{equation}
V_{ij}^{\rmout}=\frac{1}{2}\big\langle \hat{u}_{i}^{\rmout}(\infty)\hat{u}_{j}^{\rmout}(\infty) +\hat{u}_{j}^{\rmout}(\infty)\hat{u}_{i}^{\rmout}(\infty) \big\rangle,
\end{equation}
where $\hat{\boldsymbol{u}}^{\rmout} \equiv [\hat{X}_{\rmb}^{\mathrm{sel}},\hat{Y}_{\rmb}^{\mathrm{sel}},\hat{X}_{\rmc}^{\mathrm{sel}},\hat{Y}_{\rmc}^{\mathrm{sel}},\hat{X}_{\rmw}^{\mathrm{sel}},\hat{Y}_{\rmw}^{\mathrm{sel}}]^{\trp}$ is the vector formed by the field quadratures of the three output modes. The explicit expression of $\mathbf{V}^{\rmout}$ can be evaluated solving Eq.~(\ref{compact}), using the input-output relation and following the method of Ref.~\cite{Genes2008}. The resulting expressions are cumbersome and will not be reported here. Entanglement swapping with local certification can be successfully implemented when the Gaussian CV state in the hands of Alice and Bob satisfies the certifying condition of Eq.~(\ref{tcs}), and we shall investigate when this is experimentally achievable in the next Section.

%
%
\section{Results}

In order to determine the experimental conditions under which one can entangle two distant microwave modes, we consider an opto-electro-mechanical system with parameters comparable to those of recent experiments~\cite{Bochmann2013,Bagci2014,Andrews2014}. In particular our scheme is very similar to that of Ref.~\cite{Andrews2014}, where a thin vibrating SiN membrane partially coated with a small Nb electrode is coupled by radiation pressure to an optically driven Fabry-Perot cavity and capacitively coupled to a driven microwave cavity. In our case one should consider a doubly driven optical cavity. We assume a mechanical vibrational mode of the membrane with effective mass $m=10~\mathrm{ng}$, resonance frequency $\omega_{\rmm}/2\pi=10~\mathrm{MHz}$ and mechanical quality factor $Q_{\rmm}=1.5\times 10^{5}$. Moreover we have assumed the following values for the single photon optomechanical and electromechanical couplings: $g_b/ 2\pi \simeq g_c/2\pi = 152$ Hz, $g_w/2\pi = 0.266$ Hz. Two optical cavity modes are pumped by two laser beams at wavelengths $\lambda_{\rmb}=810.000~\mathrm{nm}$ and $\lambda_{\rmc}=810.328~\mathrm{nm}$, while the microwave cavity is driven by a source at wavelength $\lambda_{\rmw}=29.979~\mathrm{mm}$.

\begin{figure}[b]
\includegraphics[width=\columnwidth]{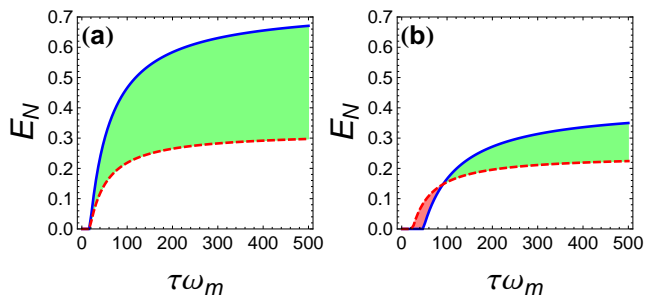}
\caption{(Color online) Log-negativity of the microwave--microwave (blue solid line) and certifying (red dashed line) entanglements as a function of the scaled output inverse bandwidth $\omega_m \tau$ for $T=50~\mathrm{mK}$ (a), and $T=100~\mathrm{mK}$ (b). The other parameters are given in the text. The parameter region corresponding to a certifying condition, i.e., a larger microwave--microwave entanglement and nonzero certifying $E_{N}$, is denoted in green (light grey), while in the red (dark grey) region certification is not possible.}
\label{tau}
\end{figure}

In order to find the best parameter region for implementing the protocol, we have studied the log negativity of the bipartite subsystem formed by the two output microwave modes, $E_{N}^{\rmw1,\rmw2}$, and of the optical certifying modes, $E_{N}^{\rmc1,\rmc2}$, as a function of the various parameters.
We have first seen that the optimal situation is when all the filtering bandwidths have the same value, i.e., $\tau_{\rmb}=\tau_{\rmc}=\tau_{\rmw}\equiv \tau$, and we have therefore restricted ourselves to this case only. Moreover, Eq.~(\ref{sympl-relat}) suggests that a higher output entanglement between the two distant microwave modes is obtained for larger optical Bell-microwave entanglement, and we can use the results of Ref.~\cite{Barzanjeh2012} for finding when this occurs. Ref.~\cite{Barzanjeh2012} shows that when the central frequency of the optical output mode is fixed at $\Omega_{\rmb/\rmc}=\mp\omega_{\rmm}$, the largest optical--microwave entanglement is achieved only around $\Omega_{\rmw}=\pm\omega_{\rmm}$ and for very narrow filtering inverse bandwidths $\tau\omega_{\rmm} \gg 1$.
In fact, optical--microwave entanglement is maximum when narrowband blue-detuned microwave and red-detuned optical output fields are selected, i.e., $\Omega_{\rmw}=\Delta_{\rmw}=-\Delta_{\rmb/\rmc}=-\Omega_{\rmb/\rmc}=\omega_{\rmm}$~\cite{Barzanjeh2012}, and we have considered such conditions in our numerical studies. Finally, one also needs to have a \textit{sufficient} entanglement between the two optical (Bell and certifying) modes, which is a necessary condition for having a detectable certifying entanglement between the two optical certifying modes, as discussed in~\cite{Abdi2012,Abdi2014}. This latter condition is achieved by choosing opposite detunings for the two optical modes, and again optimally centering the frequencies of the output modes (see for example Ref.~\cite{Genes2009}). We have chosen $\Omega_{\rmc}=\Delta_{\rmc}=-\Delta_{\rmb}=-\Omega_{\rmb}=\omega_{\rmm}$, which guarantees that the optical Bell mode
is strongly entangled with the microwave mode and also entangled with the other optical (certifying) mode.

In Fig.~\ref{tau} we plot the entanglement between two remote microwave modes as well as that between the optical certifying modes, obtained at the end of the protocol, as a function of the normalized filtering bandwidth $\tau\omega_{\rmm}$ and at two different environment temperatures.
Here we assume that the microwave cavity mode is pumped at $P_{\rmw}=35~\mathrm{mW}$, while to attain a stable system the laser powers feeding the optical modes have chosen to be $P_{\rmb}=2.0~\mathrm{mW}$ and $P_{\rmc}=2.1~\mathrm{mW}$.
We have also considered equal cavity decay rates for microwave and optical fields $\kappa_{\rmw}=\kappa_{\rmb}=\kappa_{\rmc}=0.25\omega_{\rmm}$.

It can be seen from the figure that both logarithmic negativities monotonically increase for narrower filtering bandwidths and they both tend to an asymptotic nonzero values. In Fig.~\ref{tau}(a), corresponding to $T=50$ mK, the certifying condition $E_{N}^{\rmw1,\rmw2} > E_{N}^{\rmc1,\rmc2} > 0$ is always satisfied and one can achieve significative entanglement values between the two distant microwave output modes. In Fig.~\ref{tau}(b), corresponding to $T=100$ mK, the certifying condition is not satisfied at lower values of $\tau \omega_m$, and the achievable entanglement is much lower due to the detrimental effect of thermal noise.

\begin{figure}[t]
\includegraphics[width=\columnwidth]{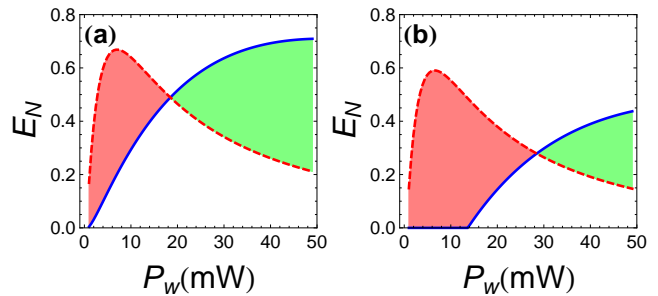}
\caption{(Color online) The microwave--microwave (blue solid line) and certifying (red dashed line) logarithmic negativities as a function of microwave input power when the optical modes are pumped at the power $P_{\rmb}=2.0~\mathrm{mW}$ and $P_{\rmc}=2.1~\mathrm{mW}$, respectively. (a) $T=50~\mathrm{mK}$ and (b) $T=100~\mathrm{mK}$. Also we set $\tau\omega_{\rmm}=500$ and the other parameters are the same as Fig.~\ref{tau}. The shading color code is the same as in Fig.~\ref{tau}.}
\label{power}
\end{figure}

In Fig.~\ref{power} we plot instead the two logarithmic negativities $E_{N}^{\rmw1,\rmw2}$ and $E_{N}^{\rmc1,\rmc2}$ versus the input power of the microwave drive at the two different temperatures $T=50$ and $T=100$ mK. The certifying condition is strongly affected by the microwave input power because the entanglement between the distant microwave modes monotonically increases with it, while the entanglement between the optical certifying modes mostly decreases with increasing power. As a consequence, the certifying condition is satisfied only at large enough input microwave power. This can be understood from the fact that larger input microwave power corresponds to a larger optical Bell-microwave mode entanglement and therefore to a smaller entanglement between the two optical modes due to entanglement monogamy, in the initial tripartite state used for the swapping protocol.

%
%
\section{Conclusion}
We have proposed a scheme based on a modified entanglement swapping protocol able to entangle two distant microwave output fields, which represent a fundamental tool for the realization of quantum networks of superconducting quantum processors. The scheme exploits the stationary multipartite entanglement that can be generated in a cryogenic quadripartite opto-electro-mechanical system in which a nanomechanical resonator is simultaneously coupled to a microwave cavity mode and to two optical cavity modes. A third party at an intermediate site performs the Bell measurement entangling the distant sites and at the same time can certify the success of the swapping protocol by determining the entanglement of two ancillary optical certifying modes. Significative entanglement between distant narrow-band microwave modes can be achieved and verified with feasible experimental setups, just thanks to the fact that the scheme does not require to send microwave modes over long distances, exposing them therefore to the detrimental effects of room temperature thermal noise.

%
%
\section*{Acknowledgments}
This work has been supported by the European Commission (ITN-Marie Curie project cQOM and FET-Open Project iQUOEMS) and by MIUR (PRIN 2011). M.A. acknowledges support from the Alexander von Humboldt Foundation.

%
%
\bibliography{swapping}

\end{document}